\documentclass[12pt]{article}
\usepackage{hyperref,amsfonts,amssymb}
\usepackage[english]{babel}

\def\dsize{\displaystyle}

\def\bq{ \begin{equation} }
\def\eq{ \end{equation} }
\def\ben{ \begin{eqnarray} }
\def\en{ \end{eqnarray} }

\def\frac#1#2{{#1\over #2}}

\def\on#1#2{\mathop{\vbox{\ialign{##\crcr\noalign{\kern2pt}
$\scriptstyle{#2}$\crcr\noalign{\kern2pt\nointerlineskip}
\kern-2pt$\hfil\displaystyle{#1}\hfil$\crcr}}}\limits}

\let\ds\displaystyle

\begin{document}
\baselineskip=15pt
\vspace{1cm} \noindent {\LARGE \textbf{On recursion operators for
elliptic models}} \vskip1cm \hfill
\begin{minipage}{13.5cm}
\baselineskip=15pt {\bf D K Demskoi ${}^{1}$,
 V V Sokolov ${}^{2}$} \\ [2ex]
{\footnotesize ${}^1$ School of Mathematics, University of New
South Wales, Sydney, NSW 2052, Australia
\\
${}^2$ Landau Institute for Theoretical Physics, Moscow, Russia
\\}

\vskip1cm{\bf Abstract.} New quasilocal recursion and Hamiltonian
operators for the Krichever-Novikov and the Landau-Lifshitz
equations are found. It is shown that the associative algebra  of
quasilocal recursion operators for these models is generated by a
couple of operators related by an elliptic curve equation.  A
theoretical explanation of this fact for the Landau-Lifshitz
equation is given in terms of multiplicators of the corresponding
Lax structure.


\end{minipage}

\vskip0.8cm \noindent{ MSC numbers: 17B80, 17B63, 32L81, 14H70 }
\vglue1cm \textbf{Address}: Landau Institute for Theoretical
Physics, Kosygina 2, 119334, Moscow, Russia

\textbf{E-mail}: sokolov@itp.ac.ru, \, demskoi@maths.unsw.edu.au \newpage

\centerline{\Large\bf Introduction}
\medskip

One of the main algebraic structures related to $1+1$-dimensional
integrable PDE of the form
\begin{equation} \label{eq1}
u_{t}=F(u, u_{x},\dots, u_n)\, , \qquad n\ge 2, \qquad u_i=
\partial^i_x u(x,t)
\end{equation}
is an infinite hierarchy of commuting flows or, the same,
(generalized) symmetries \cite{olver}
\begin{equation}\label{Geq}
u_{t_i}=G_i(u,...,u_{m_i})\, .
\end{equation}
We identify the symmetry (\ref{Geq}) with its right hand side
$G_i$. The symmetry $G$ satisfies the equation
\begin{equation} \label{def1}
D_t (G)-F_* (G)=0,
\end{equation}
where $D_t$ stands for the derivation in virtue of (\ref{eq1}) and
$F_*$ denotes the Fr\'echet derivative of $F$:
$$
F_{*}=\sum_{i=0}^n \frac{\partial^i F }{\partial u_i} D_x^i.
$$
The dual objects for symmetries are cosymmetries which satisfy the
equation
$$ \label{def2}
D_t (g)+F_*^t (g)=0,
$$
where $F_*^t$ is the differential operator adjoint to $F_*$. The
product $g\, G$ of any cosymmetry $g$ and symmetry $G$ is a total
$x$-derivative. It is well known \cite{olver} that for any conserved
density $\rho$ the variational derivative $\frac{\delta \rho}{\delta
u}$ is a cosymmetry.

The simplest symmetry for any equation (\ref{eq1}) is $u_x$. The
usual way to get other symmetries is to act to $u_x$ by a recursion
operator $\cal R$. By definition, the recursion operator is a ratio
of two differential operators that satisfies the identity
\begin{equation}\label{Lambdaeq}
[D_t-F_*,\, {\cal R}]={\cal R}_{t}-[F_{*},\,{\cal R}]=0.
\end{equation}
It follows from (\ref{def1}) and (\ref{Lambdaeq}) that for any
symmetry $G$ the expression ${\cal R} (G)$ is a symmetry as well.

Most of known recursion operators have the following special form
\begin{equation}\label{anz}
{\cal R}=R+\sum_{i=1}^k G_i D_x^{-1} g_i,
\end{equation}
where $R$ is a differential operator, $G_i$ and $g_i$ are some fixed
symmetries and cosymmetries common for all members of the hierarchy.
For all known examples the cosymmetries $g_i$ are variational
derivatives of conserved densities. Applying such operator to any
symmetry we get a local expression, (i.e. a function of finite
number of variables $u, u_x, \dots u_i, \dots$) since  the product
of any symmetry and cosymmetry belongs to $Im\, D_x$. Moreover, a
different choice of integration constants gives rise to an
additional linear combination of the symmetries $G_1,\dots, G_k$.
Probably for the first time ansatz (\ref{anz}) was used in
\cite{sokkn}. We call recursion operators (\ref{anz}) {\it
quasilocal}.

Most of known integrable equations (\ref{eq1}) can be written in a
Hamiltonian form
$$
u_t={\cal H}\left(\frac{\delta \rho}{\delta u}  \right),
$$
where $\rho$ is a conserved density and ${\cal H}$ is a
Hamiltonian operator. It is known that this operator satisfies the
equation
\begin{equation}\label{Heq}
(D_t-F_*) \, {\cal H}= {\cal H} (D_{t}+F_{*}^t),
\end{equation}
which means that $\cal H$ takes cosymmetries to symmetries.
Besides (\ref{Heq}) the Hamiltonian operator should satisfy
certain identities (see for example, \cite{olver}) equivalent to
the skew-symmetricity and the Jacobi identity for the
corresponding Poisson bracket. It is easy to see that the ratio
${\cal H}_2 {\cal H}_1^{-1}$ of any two Hamiltonian operators is a
recursion operator.

As the rule, the Hamiltonian operators are local (i.e. differential)
or quasilocal operators. The latter means that
\begin{equation}\label{anz1}
{\cal H}=H+\sum_{i=1}^m G_i D_x^{-1} \bar G_i,
\end{equation}
where $H$ is a differential operator and $G_i, \bar G_i$ are fixed
symmetries. It is clear that acting by the operator (\ref{anz1})
on any cosymmetry, we get a local symmetry.

For example, for the Korteweg-de Vries equation
$$
u_t=u_{xxx}+ 6\,u\,u_x
$$
the simplest recursion operator  (see \cite{olv1})
\begin{equation}\label{KdVr}
{\cal R}=D_x^2+4 u +2 u_x D_x^{-1}
\end{equation}
 is quasilocal with $k=1$, $G_1=2
u_x,$ and $g_1=1$. This operator is the ratio of two local
Hamiltonian operators
$$
{\cal H}_1=D_x, \qquad {\cal H}_2=D_x^3+4 u D_x+2 u_x.
$$
For the systematic investigation of quasilocal (weakly nonlocal in
terminology of \cite{malnov}) operators related to the Korteweg-de
Vries and the nonlinear Schr\-\"odinger equations see \cite{malnov,orlrub}.

It is possible to prove that for the Korteweg-de Vries equation the
associative algebra ${\bf A}$ of all quasilocal recursion operators
is generated by operator (\ref{KdVr}). In other words, this
algebra is isomorphic to the algebra of all polynomials in one
variable.

Our main observation is that it is not true for such integrable
models as the Krichever-Novikov and the Landau-Lifshitz equations.
These equations play a role of the universal models for the classes
of KdV-type and NLS-type equations. It appears that other integrable
equations from these classes are limiting cases or can be linked to
these models by differential substitutions \cite{mikh,yamil}.

It turns out that for these models, known to be elliptic, the
algebra ${\bf A}$ is isomorphic to the coordinate ring of the
elliptic curve. Since the algebra ${\bf A}$ is defined in terms of
the equation (\ref{eq1}) only, this is an invariant way to associate
a proper algebraic curve with any integrable equation and, in
particular, to give a rigorous intrinsic definition of elliptic
models.

This paper is organized as follows. In  Section {\bf 1}, we consider
the Krichever-Novikov equation \cite{krichnov}, which is the
simplest known one-field elliptic model. We show that there exist
two quasilocal recursion operators of orders 4 and 6 related by the
equation of elliptic curve. These recursion operators are ratios of
the corresponding quasilocal Hamiltonian operators. The simplest
quasilocal Hamiltonian operator of order $-1$
$$
{\cal H}_0=u_x D_x^{-1} u_x
$$
for the Krichever-Novikov equation was found in \cite{sokkn}. From
the results of this paper it follows that this equation has also a
Hamiltonian operator of order 3. In Section {\bf 1} we present one
more quasilocal Hamiltonian operator of the fifth order for the
hierarchy of the Krichever-Novikov equation. It seems to be
interesting to investigate this multi-Hamiltonian structure for the
Krichever-Novikov equation in frames of the bi-Hamiltonian approach
\cite{magri}.

In Section {\bf 2} we obtain similar results for the Landau-Lifshitz
equation. It turns out that for this model there exist quasilocal
recursion operators of orders 2 and 3 related by an elliptic curve
equation. The corresponding quasilocal Hamiltonian operators are
also found.

In Sections {\bf 1,2} we have used the direct way to construct
recursion and Hamiltonian operators based on the ansatzes
(\ref{anz}) and (\ref{anz1}). On the other hand there are several
schemes that use $L-A$-pairs for this purpose. For example, there
is a construction related to the squared eigenfunctions of the Lax
operator $L$ \cite{fok0}-\cite{for1}.

An alternative approach is based on the explicit formulas for the
$A$- operators (see \cite{sym, adl,for2}). Some version of this
approach has been suggested in \cite{GKS}. In Section {\bf 3} we
generalize the main idea of this work to the case of the
Landau-Lifshitz equation. As a result, a deep relationship between
the algebra ${\bf A}$ of quasilocal recursion operators and the
algebra of  multiplicators of the Lax structure becomes evident. In
particular, the existence of two recursion operators related by
elliptic curve equation follows from a similar property for
multiplicators.

\section{The Krichever-Novikov equation.}

The Krichever-Novikov equation \cite{krichnov, svsok} can be written
in the form
\begin{equation}
u_{t_1}=u_{xxx}-\frac{3}{2}\frac{u_{xx}^2}{u_x}+\frac{P(u)}{u_x},
\qquad P^{(V)}=0. \label{kn}
\end{equation}
Denote by $G_1$ the right hand side of (\ref{kn}).
 The fifth order symmetry of (\ref{kn}) is given by
$$
G_2=u_5-5\frac{u_4 u_2}{u_1}-\frac{5}{2}
\frac{u_3^2}{u_1}+\frac{25}{2} \frac{u_3
u_2^2}{u_1^2}-\frac{45}{8} \frac{u_2^4}{u_1^3}-\frac{5}{3} P
\frac{u_3}{u_1^2}+\frac{25}{6} P \frac{u_2^2}{u_1^3}-\frac{5}{3}
P' \frac{u_2}{u_1}-\frac{5}{18} \frac{P^2}{u_1^3}+\frac{5}{9} u_1
P''. \label{kn5}
$$
The simplest three conserved densities of (\ref{kn}) are
$$
\begin{array}{l}
\ds \rho_1=-\frac{1}{2} \frac{u_2^2}{u_1^2}-\frac{1}{3}
\frac{P}{u_1^2},\qquad
\rho_2=\frac{1}{2} \frac{u_3^2}{u_1^2}-\frac{3}{8} \frac{u_2^4}{u_1^4}+\frac{5}{6} P \frac{u_2^2}{u_1^4}+
\frac{1}{18}\frac{P^2}{u_1^4}-\frac{5}{9} P'', \\[8mm]
\ds \rho_3=\frac{u_4^2}{u_1^2}+3 \frac{u_3^3}{u_1^3}-\frac{19}{2} \frac{u_3^2 u_2^2}{u_1^4}+
\frac{7}{3}P \frac{u_3^2}{u_1^4}+\frac{35}{9} P' \frac{u_2^3}{u_1^4}+\frac{45}{8} \frac{u_2^6}{u_1^6}-
\frac{259}{36} \frac{u_2^4 P}{u_1^6}+\frac{35}{18} P^2 \frac{u_2^2}{u_1^6}\\[5mm]
\ds\qquad -\frac{14}{9} P'' \frac{u_2^2}{u_1^2}+\frac{1}{27} \frac{P^3}{u_1^6}-
\frac{14}{27} \frac{P'' P}{u_1^2}-\frac{7}{27} \frac{P'^2}{u_1^2}-\frac{14}{9} P^{(IV)} u_1^2.
\end{array}
$$

In the paper \cite{sokkn} the forth order quasilocal recursion
operator of the form
$$
{\cal R}_1=D_x^4+a_1D_x^3+a_2 D_x^2+a_3 D_x+a_4+G_1 D_x^{-1} \frac{\delta
\rho_1}{\delta u}+u_x D_x^{-1} \frac{\delta \rho_2}{\delta u},
\label{KNR1}
$$
was found. Here the coefficients $a_i$ are given by
$$
\begin{array}{l}
\ds a_1=-4 \frac{u_2}{u_1},\qquad a_2=6 \frac{u_2^2}{u_1^2}-2 \frac{u_3}{u_1}-\frac{4}{3} \frac{P}{u_1^2}, \\[5mm]
\ds a_3=-2 \frac{u_4}{u_1}+8 \frac{u_3 u_2}{u_1^2}-6 \frac{u_2^3}{u_1^3}+4 P \frac{u_2}{u_1^3}-
\frac{2}{3}\frac{P'}{u_1}, \\[5mm]
\ds a_4=\frac{u_5}{u_1}-2 \frac{u_3^2}{u_1^2}+8 \frac{u_3
u_2^2}{u_1^3}-4 \frac{u_4 u_2}{u_1^2}- 3
\frac{u_2^4}{u_1^4}+\frac{4}{9} \frac{P^2}{u_1^4}+\frac{4}{3}P
\frac{u_2^2}{u_1^4}+\frac{10}{9}P''- \frac{8}{3}
P'\frac{u_2}{u_1^2}.
\end{array}
$$
The following statement can be verified directly.

{\bf Theorem 1.} {\it There exists one more quasilocal recursion
operator for} (\ref{kn}) {\it of the form}
$$
\begin{array}{l}
\ds {\cal R}_2=D_x^6+b_1D_x^5+b_2 D_x^4+b_3 D_x^3+b_4 D_x^2+b_5 D_x+b_6-\frac{1}{2} u_x D_x^{-1} \frac{\delta \rho_3}{\delta u}\\[2mm]
\ds \qquad +G_1 D_x^{-1} \frac{\delta \rho_2}{\delta u}+G_2 D_x^{-1}
\frac{\delta \rho_1}{\delta u},
\end{array}
\label{KNR2}
$$
{\it where}
$$
\begin{array}{l}
\ds b_1=-6\frac{u_2}{u_1},\qquad b_2=-9 \frac{u_3}{u_1}-2 \frac{P}{u_1^2}+21 \frac{u_2^2}{u_1^2},\\[5mm]
\ds b_3=-11 \frac{u_4}{u_1}+60 \frac{u_3 u_2}{u_1^2}+14 P \frac{u_2}{u_1^3}-57 \frac{u_2^3}{u_1^3}-3\frac{P'}{u_1},\\[5mm]
\ds b_4=-4 \frac{u_5}{u_1}+38 \frac{u_4 u_2}{u_1^2}+22 \frac{u_3^2}{u_1^2}+99 \frac{u_2^4}{u_1^4}-155 \frac{u_3 u_2^2}{u_1^3}+\frac{34}{3} P \frac{u_3}{u_1^3}-44 P \frac{u_2^2}{u_1^4} \\[3mm]
\ds \qquad +\frac{4}{3} \frac{P^2}{u_1^4}+12 P' \frac{u_2}{u_1^2}-P'', \\[5mm]
\ds b_5=-2 \frac{u_6}{u_1}+29\frac{u_4 u_3}{u_1^2}+80 P\frac{u_2^3}{u_1^5} +
\frac{23}{3} P' \frac{u_3}{u_1^2} -104 \frac{u_2 u_3^2}{u_1^3} -70\frac{u_4 u_2^2}{u_1^3} +241 \frac{u_2^3 u_3}{u_1^4} +14 \frac{u_5 u_2}{u_1^2}\\[3mm]
\ds \qquad +\frac{20}{3}P \frac{u_4}{u_1^3}-\frac{170}{3} P\frac{u_2u_3}{u_1^4} +
\frac{4}{3} \frac{P' P}{u_1^3}-22P' \frac{u_2^2}{u_1^3} +2P'' \frac{u_2}{u_1} -
\frac{16}{3}P^2 \frac{u_2}{u_1^5} -108\frac{u_2^5}{u_1^5}, \\[5mm]
\ds b_6=\frac{u_7}{u_1}-6\frac{u_2 u_6}{u_1^2} +\frac{8}{9} P^2 \frac{u_2^2}{u_1^6}-
195 \frac{u_3^2 u_2^2}{u_1^4}+6 P\frac{u_3^2}{u_1^4}+\frac{142}{3}P \frac{u_2^4}{u_1^6}+\frac{28}{9}P' P\frac{u_2}{u_1^4} +101 \frac{u_4 u_3 u_2}{u_1^3} \\[3mm]
\ds \qquad +\frac{34}{3} P \frac{u_4 u_2}{u_1^4}-72 \frac{u_2^6}{u_1^6}-\frac{28}{9} P''' u_2+\frac{38}{3} P'' \frac{u_2^2}{u_1^2}-\frac{19}{3} P'\frac{u_4}{u_1^2}-\frac{122}{3} P' \frac{u_2^3}{u_1^4}-10 \frac{u_4^2}{u_1^2}+22 \frac{u_3^3}{u_1^3}\\[3mm]
\ds\qquad -\frac{178}{3} P \frac{u_3 u_2^2}{u_1^5}+\frac{14}{9} P^{(IV)} u_1^2+\frac{113}{3} P' \frac{u_3 u_2}{u_1^3}-\frac{2}{3} P \frac{u_5}{u_1^3}
-\frac{17}{3} P'' \frac{u_3}{u_1}-\frac{4}{3} P^2\frac{u_3}{u_1^5} -89 \frac{u_4 u_2^3}{u_1^4}\\[5mm]
\ds\qquad+236 \frac{u_3 u_2^4}{u_1^5} -13 \frac{u_5 u_3}{u_1^2}+25
\frac{u_5 u_2^2}{u_1^3}-\frac{7}{9}
\frac{P'^2}{u_1^2}-\frac{8}{27} \frac{P^3}{u_1^ 6}-\frac{4}{9}
\frac{P'' P}{u_1^2}.
\end{array}
$$
{\it The operators} ${\cal R}_1$ {\it and} ${\cal R}_2$ {\it are
related by the following elliptic curve}
\begin{equation}
{\cal R}_2^2={\cal R}_1^3-\phi {\cal R}_1-\theta, \label{rel}
\end{equation}
{\it where}
$$
\begin{array}{l}
\displaystyle \phi=\frac{16}{27}\Big((P'')^2-2 P''' P'+2 P^{(IV)} P\Big), \\[4mm]
\displaystyle \theta=\frac{128}{243}\Big(-\frac{1}{3}(P'')^3-\frac{3}{2}(P')^2P^{(IV)}+P'P''P'''
+2P^{(IV)}P''P-P(P''')^2\Big).
\end{array}
$$

{\bf Remark 1.} The relation (\ref{rel}) is understood as an
identity in the non-commutative field of pseudo-differential series
\cite{geldyk} of the form
$$\label{A}
 A=a_{m}D_x^m+a_{m-1}D_x^{m-1}+\cdots + a_0+a_{-1}D_x^{-1}+a_{-2}D_x^{-2}+\cdots\,.
$$
Here $a_i$ are local functions and the multiplication is  defined by
$$
 a D_x^k\circ b D_x^m =a\,(b D_x^{m+k}+C_k^1 D_x(b)D_x^{k+m-1}+C_k^2 D_x^2 (b)D_x^{k+m-2}+
\cdots )\, ,
$$
where $k,m\in \mathbb Z$ and
\[
C^j_n=\frac{n(n-1)(n-2)\cdots(n-j+1)}{j!}\,  .
\]

{\bf Remark 2.} It is easy to verify that $\phi$ and $\theta$ are
constants for any polynomial $P(u)$, where $\hbox{deg}\,P\le 4$.
Under M\"obius transformations of the form
$$
u\rightarrow \frac{\alpha u+\beta}{\gamma u+\delta}
$$
in equation (\ref{kn}) the polynomial $P(u)$ changes according
to the same rule as in the differential $\displaystyle \omega =
\frac{d u}{\sqrt{P(u)}}.$ The expressions  $\phi$ and $\theta$ are
invariants with respect to the M\"obius group action.

{\bf Remark 3.} Of course, the ratio ${\cal R}_3={\cal R}_2 {\cal
R}_1^{-1}$ satisfies equation (\ref{Lambdaeq}). It belongs to the
noncommutative field of differential operator fractions \cite{ore}.
Any element of this field can be written in the form $P_1 P_2^{-1}$
for some differential operators $P_i$. So, according to our
definition, ${\cal R}_3$ is a recursion operator of order 2.
However, this operator is not quasilocal and it is unclear how to
apply it even to the simplest commuting flow $u_x$.

The recursion operators presented above appear to be ratios
$$
{\cal R}_1={\cal H}_1 {\cal H}_0^{-1}, \qquad  {\cal R}_2={\cal H}_2
{\cal H}_0^{-1}
$$
of the following quasilocal Hamiltonian operators
$$
\begin{array}{l}
{\cal H}_1= \frac{1}{2}(u_x^2 D_x^3+D_x^3 u_x^2)+(2 u_{xxx} u_x-\frac{9}{2} u_{xx}^2-\frac{2}{3} P)D_x+D_x(2 u_{xxx} u_x-\frac{9}{2} u_{xx}^2-\frac{2}{3} P) \\[2mm]
\qquad +G_1 D_x^{-1} G_1+  u_x D_x^{-1} G_2 +G_2 D_x^{-1} u_x,
\label{KNH1}
\end{array}
$$
$$
\begin{array}{l}
{\cal H}_2= \frac{1}{2}(u_x^2 D_x^5+D_x^5 u_x^2)+(3u_{xxx} u_x-\frac{19}{2} u_{xx}^2-P)D_x^3+D_x^3(3 u_{xxx} u_x-\frac{19}{2} u_{xx}^2-P)\\[2mm]
\qquad +h D_x+D_x h+G_1 D_x^{-1} G_2+G_2 D_x^{-1} G_1+u_x D_x^{-1} G_3 +G_3 D_x^{-1} u_x,
\label{KNH2}
\end{array}
$$
where
$$
h=u_{xxxxx} u_x-9 u_{xxxx} u_{xx}+\frac{19}{2} u_{xxx}^2-\frac{2}{3} \frac{u_{xxx}}{u_x} (5 P-39 u_{xx}^2)+\frac{u_{xx}^2}{u_x^2} (5 P-9 u_{xx}^2)+\frac{2}{3} \frac{P^2}{u_x^2}+u_x^2 P'',
$$
and $G_3={\cal R}_1(G_1)={\cal R}_2 (u_x)$ is the seventh order
symmetry of (\ref{kn}):
$$
\label{KNG7}
\begin{array}{l}
\dsize G_3=u_7-7 \frac{u_2 u_6}{u_1}-\frac{7}{6} \frac{u_5}{u_1^2}(2 P+12 u_3 u_1-27 u_2^2)-\frac{21}{2} \frac{u_4^2}{u_1}+\frac{21}{2} \frac{u_4}{u_1^3} u_2(2 P-11 u_2^2)\\[2mm]
\dsize\qquad -\frac{7}{3} \frac{u_4}{u_1^2}(2 P' u_1-51 u_2 u_3)  +\frac{49}{2} \frac{u_3^3}{u_1^2}+\frac{7}{12} \frac{u_3^2}{u_1^3} (22 P-417 u_2^2)+\frac{2499}{8} \frac{u_2^4}{u_1^4} u_3\\[2mm]
\dsize\qquad  +\frac{91}{3} P'\frac{u_2}{u_1^2} u_3-\frac{595}{6} P \frac{u_2^2}{u_1^4} u_3-\frac{35}{18} \frac{u_3}{u_1^4} (2 P'' u_1^4-P^2)-\frac{1575}{16} \frac{u_2^6}{u_1^5}+\frac{1813}{24} \frac{u_2^4}{u_1^5} P\\[2mm]
\dsize\qquad -\frac{203}{6}\frac{u_2^3}{u_1^3} P' +\frac{49}{36} \frac{u_2^2}{u_1^5} (6 P'' u_1^4-5 P^2)-\frac{7}{9} \frac{u_2}{u_1^3} (2 P''' u_1^4-5 P P')+\frac{7}{54} \frac{P^3}{u_1^5}\\[2mm]
\dsize\qquad -\frac{7}{9} P'' \frac{P}{u_1}+\frac{7}{9} P'''' u_1^3-\frac{7}{18} \frac{P'^2}{u_1}.
\end{array}
$$

\section{The Landau-Lifshitz equation.}

In this section we consider the Landau-Lifshitz equation written in
the form
\begin{equation}
\begin{array}{l}
u_{t_2}=-u_{xx}+2\psi\, (u_x^2-P(u) )+\frac{1}{2}P'(u) \\[2mm]
v_{t_2}=\quad v_{xx}+2\psi\, (v_x^2-P(v))-\frac{1}{2}P'(v),
\label{LL2}\end{array}
\end{equation}
where
$$\psi=(u-v)^{-1},$$
and  $P$ is an arbitrary fourth degree polynomial. The usual
vectorial form of Landau-Lifshitz equation gives rise to a system of
the form (\ref{LL2}) after the stereographic projection (see Section
{\bf 3} for details).

The third order symmetry of (\ref{LL2}) is given by
$$
\begin{array}{l}
u_{t_3}=u_{xxx}-6 u_x u_{xx} \psi+6 u_x^3\psi^2-\frac{1}{2} u_x P''(v)-3 u_x \psi P'(v)-6 u_x \psi^2 P(v), \\[2mm]
v_{t_3}=v_{xxx}+6 v_x v_{xx} \psi+6 v_x^3\psi^2-\frac{1}{2} v_x
P''(v)-3 v_x \psi P'(v)-6 v_x \psi^2 P(v).
\end{array}
\label{LL3}
$$
In case of the system of evolution equation (\ref{LL2}) the
symmetries and cosymmetries can be treated as two-dimensional
vectors. We introduce the following notation for symmetries:
\begin{equation}
G_1=(u_x, v_x)^t, \qquad  G_2=(u_{t_2},v_{t_2})^t, \qquad
G_3=(u_{t_3},v_{t_3})^t. \label{SS}
\end{equation}
Thus $G_{ij}$ is $j$-th component of the symmetry $G_i$. The
simplest cosymmetries are given by
\begin{equation}
g_1=\psi^2(v_x,-u_x)^t,\qquad  g_2=\psi^2(v_{t_2},-u_{t_2})^t,
\qquad  g_3=\psi^2(v_{t_3},-u_{t_3})^t. \label{GG}
\end{equation}
These cosymmetries are variational derivatives $(\frac{\delta
\rho}{\delta u}, \frac{\delta \rho}{\delta v})^t$ of the following
conserved densities
$$
\begin{array}{l}
\rho_1=\frac{1}{2}\psi (u_x+v_x)  , \qquad \rho_2=\psi^2(u_xv_x-P(v))-\frac{1}{2}\psi P'(v)-\frac{1}{12} P''(v)  ,\\[2mm]
\rho_3=\frac{1}{2}\psi^2( u_x v_{xx} - u_{xx} v_x)+\psi^3 u_x v_x(u_x+v_x)-\frac{1}{2} u_x\psi (4 \psi^2 P(u)+ P''(u)-3 \psi P'(u)) .
\end{array}
$$
The coefficients of operators $F_{*}, {\cal R}$, and ${\cal H}$ from
(\ref{Lambdaeq}), (\ref{Heq}) are matrices. For the quasilocal
recursion operators the non-local terms can be written as $A
D_x^{-1} B^t,$ where $A$ and $B$ are  $k\times 2$ matrices whose
columns are symmetries and cosymmetries, correspondingly. For
quasilocal Hamiltonian operators the columns are symmetries for both
$A$ and $B$.

{\bf Theorem 2.} {\it Equation} (\ref{LL2}) {\it possesses the
following quasilocal  recursion operators}:
\begin{equation}
\begin{array}{l}
{\cal R}_1= \left(
\begin{array}{cc}
R^1_{11} & 0 \\[2mm]
0 & R^1_{22}
\end{array}\right)-2 \left(
\begin{array}{cc}
G_{11} & G_{21} \\
G_{12} & G_{22}
\end{array}\right)D_{x}^{-1} \left(
\begin{array}{cc}
g_{21} & g_{22} \\
g_{11} & g_{12}
\end{array}\right)
\end{array}
\end{equation}
{\it and}
\begin{equation}
\begin{array}{l}
{\cal R}_2 = \left(
\begin{array}{cc}
R^2_{11} & R^2_{12} \\[2mm]
R^2_{21} & R^2_{22}
\end{array}\right)+2 \left(
\begin{array}{ccc}
G_{11} & G_{21}& G_{31} \\
G_{12} & G_{22}& G_{32}
\end{array}\right)D_{x}^{-1} \left(
\begin{array}{cc}
g_{31} & g_{32} \\
g_{21} & g_{22} \\
g_{11} & g_{12}
\end{array}\right),
\end{array}
\end{equation}
{\it where} $G_{ij}$ {\it and} $g_{ij}$ {\it are defined by}
(\ref{SS}), (\ref{GG}), {\it and}
$$
\begin{array}{l}
R^1_{11}=D_{x}^2-4\psi u_x D_{x}+2\psi^2 u_x(v_x +3 u_x)-2 \psi u_{xx}-\frac{1}{3} P''(v)-4 P(v)\psi^2-2 \psi P'(v), \\[3mm]
R^1_{22}=D_{x}^2+4\psi v_x D_{x}+2\psi^2 v_x( u_x+3 v_x)+2 \psi v_{xx}-\frac{1}{3} P''(v)-4 P(v)\psi^2-2 \psi P'(v), \\[3mm]
R^2_{11}=D_{x}^3-6\psi u_x D_{x}^2+\big(6\psi^2(3 u_x^2- P(v))-6
\psi u_{xx}-\frac{1}{2} P''(v)-3 \psi P'(v)\big)D_{x}+\psi( P''(v)
u_x-2 u_{xxx}) \\[2mm]
\qquad\ \  +4\psi^3 u_x(6 P(v)-6 u_x^2-v_x^2-u_x v_x)+\psi^2(2 v_x u_{xx}-2 u_x v_{xx}+9 P'(v) u_x+18 u_x u_{xx}), \\[3mm]
R^2_{12}=2\psi^2
(P(u)-u_x^2)D_{x}-4\psi^3\big(P(u)(u_x-v_x)+u_x^2v_x-u_x^3\big)+\psi^2u_x(P'(u)-2 u_{xx}),\\[3mm]
R^2_{21}=-2\psi^2 (P(v)-v_x^2)D_{x}+4\psi^3\big(P(v)( u_x-v_x)-v_x^2 u_x+ v_x^3\big)-\psi^2 v_x(P'(v)-2 v_{xx}), \\[3mm]
R^2_{22}=-D_{x}^3-6 v_x\psi D_{x}^2+(6\psi^2(P(v)-3 v_x^2 )-6 v_{xx} \psi+3 \psi P'(v)+\frac{1}{2} P''(v))D_{x} \\[2mm]
\qquad \ \ -4\psi^3v_x(6 v_x^2+v_x  u_x+u_x^2)+2\psi^2(3 v_x
P'(v) -9 v_x v_{xx}-u_x v_{xx}+v_x u_{xx}-\frac{3}{2} v_x P'(u)) \\[2mm]
\qquad \ \ +12\psi^3v_x( P(v)+P(u))+\psi(v_x P''(v)-2 v_{xxx}).
\end{array}
$$
{\it Operators} ${\cal R}_1$ {\it and} ${\cal R}_2$ {\it are related
by the following elliptic curve equation}
\begin{equation}\label{curv1}
{\cal R}_2^2-{\cal R}_1^3-\varphi {\cal R}_1-\vartheta E=0,
\end{equation}
{\it where} $E$ {\it stands for the unity matrix, and}
$$
\begin{array}{l}
\varphi=12\psi^4P(v)\big(P(u)- P(v)\big)+4 \psi^3\big( P'(v)P(u)-P(v) P'(u)-3P(v) P'(v)\big)\\[3mm]
\qquad -\psi^2(2 P''(v) P(v)+P'(v) P'(u)+3 P'(v)^2)-\psi P''(v) P'(v)-\frac{1}{12} P''(v)^2,\\[4mm]
\vartheta= 16\psi^6 P(v)\big(P(v)-P(u)\big)^2+8\psi^5\big(P(v)- P(u)\big)\big(P(v)P'(u)+3 P'(v)P(v)\big)\\[3mm]
\qquad +\psi^4 \big(P(v)P'(u)^2-3 P'(v)^2 P(u)+6 P(v)\big(P'(u)P'(v)+2 P'(v)^2\big)+4 P(v)P''(v)\big(P(v)-P(u)\big)\big) \\[3mm]
\qquad +\psi^2P''(v)\big(P'(v)^2+\frac{1}{3} P''(v) P(v)+\frac{1}{6} P'(v)P'(u)\big)+\frac{1}{6}\psi P''(v)^2 P'(v)+\frac{1}{108} P''(v)^3\\[3mm]
\qquad +\psi^3(P'(v)^2 P'(u)+\frac{2}{3}P''(v) \big(P(v)P'(u)-P'(v) P(u)\big)+4 P''(v) P(v)P'(v)+2 P'(v)^3). \\[3mm]
\end{array}
$$

It is easy to verify that $\varphi$ and $\vartheta$ are constants
for any polynomial $P$, if $\hbox{deg}\,P\le 4$. They are invariants
with respect to the M\"obius group action just as in the case of the
Krichever-Novikov equation (see Remark 2).

{\bf Remark 4.} Possibly the odd recursion operator found in
\cite{yan} is a rational function of recursion operators from
Theorem {\bf 2}.

It turns out that the above recursion operators are ratios
$$
{\cal R}_1={\cal H}_1 {\cal H}_0^{-1}, \qquad  {\cal R}_2={\cal H}_2
{\cal H}_0^{-1}
$$
of the following quasilocal Hamiltonian operators
$$\label{LLHAM1}
{\cal H}_0=\left(
\begin{array}{cc}
0&\frac{1}{\psi^2} \\
-\frac{1}{\psi^2} & 0 \label{J0}\end{array}\right),
$$
$$
\begin{array}{l}
{\cal H}_1= \left(
\begin{array}{cc}
0 & {\cal H}^1_{12} \\
{\cal H}^1_{21} & 0
\end{array}\right)-2\left(
\begin{array}{cc}
G_{11} & G_{21} \\
G_{12} & G_{22}
\end{array}\right)D_{x}^{-1}\left(
\begin{array}{cc}
G_{21} & G_{22} \\
G_{11} & G_{12}
\end{array}\right),
\end{array}\label{LLHAM2}
$$
$$
\begin{array}{l}
{\cal H}_2= \left(
\begin{array}{cc}
{\cal H}^2_{11} & {\cal H}^2_{12} \\
{\cal H}^2_{21} & {\cal H}^2_{22}
\end{array}\right)+2\left(
\begin{array}{ccc}
G_{11} & G_{21} & G_{31} \\
G_{12} & G_{22} & G_{32}
\end{array}\right)D_{x}^{-1}\left(
\begin{array}{cc}
G_{31} & G_{32} \\
G_{21} & G_{22} \\
G_{11} & G_{12}
\end{array}\right),
\end{array}
\label{LLHAM3}
$$
where
$$
\begin{array}{l}
 \dsize {\cal H}^1_{12}=\frac{1}{\psi^2}D_{x}^2-4\frac{v_x}{\psi}D_{x}-2 \frac{v_{xx}}{\psi}+2 v_x^2+6 u_x v_x-
 \frac{1}{3} \frac{P''(v)}{\psi^2}-2 \frac{P'(v)}{\psi}-4 P(v), \\[4mm]
\dsize {\cal H}^1_{21}=-\frac{1}{\psi^2}D_{x}^2-4\frac{u_x}{\psi} D_{x}-2
\frac{u_{xx}}{\psi}-6 u_x v_x-2 u_x^2+
 \frac{1}{3} \frac{P''(v)}{\psi^2}+2 \frac{P'(v)}{\psi}+4 P(v), \\[4mm]
\dsize {\cal H}^2_{11}=2 (u_x^2-P(u))D_{x}+2 u_x u_{xx}-P'(u) u_x, \qquad {\cal H}^2_{22}=2 (v_x^2-P(v))D_{x}+2 v_x v_{xx}-P'(v) v_x,
\\[4mm]
\dsize {\cal H}^2_{12}= \frac{1}{\psi^2} D_{x}^3-6
\frac{v_x}{\psi}D_{x}^2+\left(12 u_x v_x+6\bigg(
v_x^2-\frac{v_{xx}}{\psi}-P(v)\bigg)
-\frac{1}{2}\frac{P''(v)}{\psi^2}-3  \frac{P'(v)}{\psi}\right)D_{x}+v_x \frac{P''(v)}{\psi}\\[3mm]
\dsize \qquad\ \ -2 \frac{v_{xxx}}{\psi}+6
v_x v_{xx}+8
v_x u_{xx}+4 u_x v_{xx}+4 \psi(u_x+v_x)(3 P(v)-4 u_x v_x)+P'(v)(3 u_x+6 v_x ),\\[4mm]
\dsize {\cal H}^2_{21}=\frac{1}{\psi^2} D_{x}^3+6
\frac{u_x}{\psi}D_{x}^2+\left(12 u_x v_x +6\bigg(u_x^2+
\frac{u_{xx}}{\psi}-P(v)\bigg)-\frac{1}{2} \frac{P''(v)}{\psi^2}-3 \frac{P'(v)}{\psi}\right)D_{x} \\[3mm]
\dsize \qquad\ \ +3 v_x P'(u)+2 \frac{u_{xxx}}{\psi}-u_x\frac{P''(v)}{\psi}+6
u_x u_{xx}+8 u_x v_{xx}+4 v_x u_{xx}-6 P'(v) u_x\\[3mm]
\qquad\qquad\ \ +16 u_x v_x (u_x+v_x)\psi-12\psi (P(v) u_x+P(u) v_x).
\end{array}
$$

\section{Recursion operators and multiplicators.}

The original vector form of Landau-Lifshitz reads as follows
\begin{equation}\label{LL}
{\bf U}_{t}={\bf U}\times {\bf U}_{xx}+ {\bf U} \times J{\bf U}.
\end{equation}
Here ${\bf U}=(u_{1},u_{2},u_{3})$, $\vert {\bf U} \vert=1$, symbol
$\times$ stands for the vector product, and $J=\mbox{diag} (p,q,r)$
is an arbitrary constant diagonal matrix.
The usual way to represent equation (\ref{LL}) as a two-component
system is to make use of the stereographic projection. The
transformation
\begin{equation}\label{transf}
{\bf U}=(1-uv,i+i u v,u+v)\, \psi, \quad \mbox{where} \quad i^2=-1,
\end{equation}
coincides with the stereographic projection if we set $v=-1/\bar u$.
This transformation takes equation (\ref{LL}) into the system of the
form (\ref{LL2}):
\begin{equation}
\label{LL2S}
\begin{array}{l}
u_\tau=-u_{xx}+2\psi (u_x^2-P(u))+\frac{1}{2}P'(u),\\[2mm]
v_\tau=v_{xx}+2\psi (v_x^2-P(v))-\frac{1}{2} P'(v),
\end{array}
\end{equation}
where $t=-i \tau$,
$$P(u)=\frac{1}{4}(p-q)u^4 -\frac{1}{2}(p+q-2 r)u^2+\frac{1}{4}(p-q).$$
Here and below the expression $({\bf A}, {\bf B})$ denotes the
standard Euclidean scalar product of the vectors {\bf A} and {\bf
B}.
Note that for equation (\ref{LL2S}) curve (\ref{curv1}) has the
form
$$
{\cal R}_2^2=\Big({\cal R}_1+\frac{2 p-q-r}{3} E\Big) \Big({\cal
R}_1+\frac{2 q-p-r}{3} E \Big) \Big({\cal R}_1+\frac{2 r-p-q}{3}
E\Big).
$$

Recall the algebraic structure lying behind the elliptic Lax pair
\cite{sklyan} for the Landau-Lifshitz equation. Let
$$
{\bf e_{1}}=\pmatrix{0&0&1 \cr 0&0&0 \cr -1&0&0\cr}, \quad {\bf
e_{2}}=\pmatrix{0&0&0 \cr 0&0&1 \cr 0&-1&0\cr}, \quad {\bf
e_{3}}=\pmatrix{0&1&0 \cr -1&0&0 \cr 0&0&0\cr}
$$
be the standard basis in the Lie algebra $so(3)$. The Lax operator
$L$ for (\ref{LL}) is given by
\begin{equation}\label{LAX}
L=D_x-\sum_{j=1}^3 u_{i} {\bf E_{i}},
\end{equation}
where
$$
{\bf E_{1}}=\frac{1}{\lambda}\,{\bf e_{1}} \sqrt{1-p \lambda^{2}},
\qquad {\bf E_{2}}=\frac{1}{\lambda}\,{\bf e_{2}} \sqrt{1-q
\lambda^{2}}, \qquad {\bf E_{3}}=\frac{1}{\lambda}\,{\bf e_{3}}
\sqrt{1-r \lambda^{2}}.
$$
The operators $A_i$ defining the Lax representations
\begin{equation}\label{Laxh}
L_{t_{i}}=[A_i,\,L]
\end{equation}
for the commuting flows $${\bf U}_{t_{i}}={\bf H}^{(i)} ({\bf
U},{\bf U}_x,\dots)
$$ from the Landau-Lifshitz hierarchy belong to the Lie algebra
${\cal G}$ generated by ${\bf E_{1}},{\bf E_{2}},{\bf E_{3}}.$

{\bf Lemma 1.} {\it The algebra} ${\cal G}$ {\it is spanned by}
\begin{equation}\label{basis} \frac{1}{\lambda^{2i}}{\bf E_{1}},\quad \frac{1}{\lambda^{2i}}{\bf
E_{2}}, \quad \frac{1}{\lambda^{2i}}{\bf E_{3}},\quad
\frac{1}{\lambda^{2i}} {\bf \bar E_{1}},\quad
\frac{1}{\lambda^{2i}}{\bf \bar E_{2}},\quad
\frac{1}{\lambda^{2i}}{\bf \bar E_{3}}, \qquad i=0,1,2,\dots\,
,\end{equation}
{\it where}
$$
{\bf \bar E_{1}}=\frac{1}{\lambda^2}\,{\bf e_{1}} \sqrt{1-q
\lambda^{2}} \sqrt{1-r \lambda^{2}}, \quad {\bf \bar
E_{2}}=\frac{1}{\lambda^2}\,{\bf e_{2}} \sqrt{1-p \lambda^{2}}
\sqrt{1-r \lambda^{2}}, \quad {\bf \bar
E_{3}}=\frac{1}{\lambda^2}\,{\bf e_{3}} \sqrt{1-p
\lambda^{2}}\sqrt{1-q \lambda^{2}}.
$$

Our main observation is that the recursion operators are in one-to
one correspondence with the multiplicators (see \cite{ostap,golsok})
of the algebra ${\cal G}.$

{\bf Definition.} A (scalar) function $\mu(\lambda)$ is called the
${\it multiplicator}$ for the algebra ${\cal G}$ if $\mu(\lambda)
\,{\cal G} \subset {\cal G}.$ The order of pole of $\mu(\lambda)$ at
$\lambda=0$ is called the order of the multiplicator.

It is easy to prove the following

{\bf Lemma 2.} {\it The set of all multiplicators for} ${\cal G}$
{\it coincides with the polynomial ring generated by} $1$,
$$
\mu_1(\lambda)=\frac{1}{\lambda^2}, \qquad \mu_2(\lambda)=
\frac{\sqrt{1-p \lambda^{2}}\sqrt{1-q \lambda^{2}}\sqrt{1-r
\lambda^{2}}}{\lambda^3}.
$$

It is clear that
$$
\mu_2^2=(\mu_1 -p) (\mu_1 -q)(\mu_1-r).
$$
Thus the ring of multiplicators is isomorphic to the coordinate ring
of an elliptic curve.

The following construction establishes a correspondence between
multiplicators and recursion operators. Let $\mu$ be a
multiplicator of order $k>0$. To find a relation between operators
$A_{n}$ and $A_{n+k}$ we use the following anzats
$$
A_{n+k}=\mu A_n+R_n, \qquad R_n\in {\cal G}, \qquad
\mbox{ord}\,R_n<k.
$$
Substituting this into (\ref{Laxh}) and taking into account
(\ref{LAX}), we get
\begin{equation}\label{rel1}
\sum_{j=1}^3 H_{j}^{(n+k)} {\bf E_{j}}= \mu \sum_{j=1}^3
H_{j}^{(n)} {\bf E_{j}}+\Big[\sum_{j=1}^3 u_{j} {\bf E_{j}} ,\,
R_n\Big]-\frac{d R_{n}}{dx }.
\end{equation}
Both sides of this relation belong to ${\cal G}.$ Equating in
(\ref{rel1}) the coefficients of basis elements (\ref{basis}), we
find step by step unknown coefficients of $R_n$ and eventually an
expression for $H_{j}^{(n+k)}$ in terms of $H_{j}^{(n)}$ i.e. a
recursion operator of order $k$.

For the simplest multiplicator $\mu_1=\lambda^{-2},$ we have
$$
R_n=\sum_{j=1}^3 M_j {\bf E}_j +\sum_{j=1}^3 F_j {\bf \bar E}_j .
$$
It is easy to verify that (\ref{rel1}) is equivalent to the
following identities:
$$
\begin{array}{l}
\mbox{1.}\ \ {\bf H}^{(n)}-{\bf F}\times {\bf U}=0, \\[3mm]
\mbox{2.}\ \ {\bf F}_x+{\bf U}\times {\bf M}=0, \\[3mm]
\mbox{3.}\ \ {\bf H}^{(n+2)}={\bf M}_x+{\bf F}\times J{\bf U},
\end{array}
\label{e1}
$$
where ${\bf M}=(M_1,M_2,M_3)$, ${\bf F}=(F_1,F_2,F_3).$ It follows
from the first identity and the condition $\vert {\bf U} \vert=1$
that
\begin{equation}\label{FF}
{\bf F}={\bf U}\times {\bf H}^{(n)}+f\, {\bf U}.
\end{equation}
To find $f$ we note that the second identity implies  $({\bf U},
{\bf F}_x)=0.$ Substituting the expression (\ref{FF}) to this
relation, we get
$$
f=D_x^{-1}({\bf U}, {\bf H}^{(n)}\times {\bf U}_x).
$$
The second identity can be rewritten as
$$
{\bf M}={\bf U}\times {\bf F}_x+m {\bf U}={\bf U}({\bf U},{\bf
H}^{(n)}_x)- {\bf H}^{(n)}_x+f{\bf U}\times {\bf U}_x+m {\bf U}
$$
for some function $m$. To find $m,$ we substitute into the third
identity the latter expression for $\bf M$ and take the scalar
product by $\bf U$ of both sides of the relation thus obtained.
Taking into account that $({\bf H}^{(n+2)}, {\bf U})=0$, we get
$$
m=D_{x}^{-1}\Big((J{\bf U},{\bf H}^{(n)})-({\bf U}_x,{\bf
H}^{(n)}_x)\Big).
$$
Now the third identity produces the following recursion operator for
equation (\ref{LL}):
\begin{equation}
\begin{array}{l}
{\bf H}^{(n+2)}=({\bf U},{\bf H}^{(n)}_{xx}){\bf U}-{\bf H}^{(n)}_{xx}+({\bf U},{\bf H}^{(n)}_x){\bf U}_x-({\bf U},{\bf U}_x\times {\bf H}^{(n)}){\bf U}\times {\bf U}_x+({\bf U},J{\bf U}){\bf H}^{(n)}\\[2mm]
\quad -({\bf U}\times {\bf U}_{xx}+{\bf U}\times J{\bf
U})D_x^{-1}({\bf U}, {\bf U}_x\times {\bf H}^{(n)} )-{\bf
U}_xD_{x}^{-1}\left(({\bf U}_x,{\bf H}^{(n)}_x)-(J{\bf U},{\bf
H}^{(n)})\right).
\end{array}
\label{R11}
\end{equation}
The recursion operator (\ref{R11}), along with the bi-Hamiltonian
structure of the Landau-Lifshitz equation, was originally discovered
in \cite{fokas}, whereas explicit formulas for higher symmetries
were given earlier by Fuchssteiner \cite{fuch}.

For the second multiplicator $\mu_2$ we set
\begin{equation}
R_n=\sum_{j=1}^3 K_j {\bf E}_j +\sum_{j=1}^3 M_j {\bf \bar E}_j+ \frac{1}{\lambda^2}\sum_{j=1}^3 F_j {\bf E}_j.
\label{mu2}
\end{equation}
Upon the substitution of (\ref{mu2}) to (\ref{rel1}) we get the
following identities
\begin{equation}
\begin{array}{l}
\mbox{1.}\ \ {\bf H}^{(n)}={\bf F}\times {\bf U}, \\ [2mm]
\mbox{2.}\ \ {\bf F}_x={\bf M}\times {\bf U}, \\ [2mm]
\mbox{3.}\ \ {\bf M}_x-J{\bf H}^{(n)}={\bf K}\times {\bf U}, \\ [2mm]
\mbox{4.}\ \ {\bf H}^{(n+3)}={\bf K}_x+{\bf M}\times J{\bf U}.
\end{array}
\label{e2}
\end{equation}
The first two relations in (\ref{e2}) are analogous to (\ref{e1}), and therefore
$$
{\bf F}={\bf U}\times {\bf H}^{(n)}+f {\bf U}, \qquad
f=D_{x}^{-1}({\bf U},{\bf H}^{(n)}\times {\bf U}_x),
$$
$$
{\bf M}=({\bf U},{\bf H}^{(n)}_x){\bf U}-{\bf H}^{(n)}_x+f {\bf U}\times {\bf U}_x +m {\bf U}.
$$
It follows from $(\ref{e2})_3$ that
$$
\begin{array}{l}
{\bf K}={\bf U}\times ({\bf M}_x-J{\bf H}^{(n)})+k{\bf U}= ({\bf U},{\bf H}^{(n)}_x){\bf U}\times {\bf U}_x-{\bf U}\times J{\bf H}^{(n)}-{\bf U}\times{\bf H}^{(n)}_{xx}\\[2mm]
\qquad -{\bf U}_x({\bf U},{\bf H}^{(n)}\times {\bf U}_x)-(({\bf U}_x,{\bf U}_x){\bf U}+{\bf U}_{xx})f+m{\bf U}\times {\bf U}_x+\kappa {\bf U}.
\end{array}
$$
Functions $m$ and $\kappa$ can be found from the conditions
$$({\bf U}, {\bf M}_x-J{\bf H}^{(n)})=0, \qquad ({\bf U}, {\bf H}^{(n+3)})=0,$$
which yield
$$
\begin{array}{l}
m=D_{x}^{-1}(({\bf U},J{\bf H}^{(n)})-({\bf U}_x,{\bf H}_x^{(n)})) , \\[2mm]
\kappa=
D_{x}^{-1}( ({\bf U},  {\bf U}_x\times J {\bf H}^{(n)})-({\bf U},{\bf H}^{(n)}_{xx}\times {\bf U}_x)
 -\frac{1}{2}({\bf U},{\bf H}^{(n)}\times {\bf U}_x)(({\bf U},J{\bf U})+({\bf U}_x,{\bf U}_x)) \\[2mm]
\qquad
+({\bf U},{\bf H}^{(n)}_x\times J {\bf U}))-\frac{1}{2}(({\bf U}_x,{\bf U}_x)-({\bf U},J{\bf U}))D_{x}^{-1}({\bf U},{\bf H}^{(n)}\times {\bf U}_x).
\end{array}
$$
Thus the second recursion operator reads as
\begin{equation}\label{R12}
\begin{array}{l}
{\bf H}^{(n+3)}={\bf H}^{(n)}_{xxx} \times {\bf U}+{\bf H}^{(n)}_{xx}\times {\bf U}_x+ ({\bf H}^{(n)}_{xx},{\bf U}){\bf U}\times {\bf U}_x- ({\bf U},{\bf H}^{(n)}_{xx}\times {\bf U}_x){\bf U} \\[3mm]
\qquad
+({\bf H}^{(n)}_x,{\bf U}) {\bf U}\times J{\bf U}+({\bf H}^{(n)}_x,{\bf U}){\bf U}\times {\bf U}_{xx}-{\bf H}^{(n)}_x\times J{\bf U}+({\bf U},{\bf H}^{(n)}_x\times J{\bf U}){\bf U}\\[3mm]
\qquad
-({\bf U},{\bf H}^{(n)}_x\times {\bf U}_x){\bf U}_x
-2 ({\bf U},{\bf H}^{(n)}\times {\bf U}_x)({\bf U}_x,{\bf U}_x){\bf U}-{\bf U}_x\times J{\bf H}^{(n)}-{\bf U}\times J{\bf H}^{(n)}_x
\\[3mm] \qquad
+({\bf U},{\bf U}_x\times J{\bf H}^{(n)}){\bf U}+ (J{\bf H}^{(n)},{\bf U}){\bf U}\times {\bf U}_x- ({\bf U},{\bf H}^{(n)}\times {\bf U}_{xx}){\bf U}_x\\[3mm]
\qquad -2 ({\bf U},{\bf H}^{(n)}\times {\bf U}_x){\bf U}_{xx}
-({\bf U}\times {\bf U}_{xx}+{\bf U}\times J{\bf
U})D_x^{-1}\left(({\bf H}^{(n)}_x,{\bf U}_x)-(J{\bf H}^{(n)},{\bf
U})\right)
\\[3mm]\qquad
-\left( ({\bf U}_{xx}+\frac{3}{2} {\bf U}({\bf U}_x,{\bf U}_x))_x-\frac{3}{2} {\bf U}_x({\bf U},J{\bf U})\right)D^{-1}_x({\bf U},{\bf H}^{(n)}\times {\bf U}_x) \\[3mm] \qquad
+{\bf U}_x D_x^{-1}\Big(
({\bf U},{\bf U}_x\times J{\bf H}^{(n)})-({\bf U},{\bf H}^{(n)}_{xx}\times {\bf U}_x)-\frac{1}{2}({\bf U},{\bf H}^{(n)}\times {\bf U}_x) ({\bf U},J{\bf U})\\[3mm]
\qquad\qquad -\frac{1}{2} ({\bf U},{\bf H}^{(n)}\times {\bf U}_x) ({\bf U}_x,{\bf U}_x)+({\bf H}^{(n)}_x\times J{\bf U},{\bf U})
\Big).
\end{array}
\end{equation}

One can verify that under transformation (\ref{transf}) the
operators defined by formulas (\ref{R11}) and (\ref{R12}) become
$\frac{1}{3}(p+q+r)E-{\cal R}_1$ and ${\cal R}_2$ correspondingly.

\vskip.3cm \noindent {\bf Acknowledgments.} The authors are grateful
to Professors I.M.~Krichever and W.K.~Schief for useful discussions.
The research was partially supported by: RFBR grant 05-01-00775, NSh
grants 1716.2003.1 and 2044.2003.2.

\end{document}